\documentclass{Interspeech}

\interspeechcameraready

\usepackage{subcaption}
\usepackage{caption} 
\usepackage{stfloats} 
\usepackage{arydshln}

\newcommand{\lt}{\ensuremath <}

\title{Neural Speech Extraction with Human Feedback}

\author[affiliation={1}]{Malek}{Itani}
\author[affiliation={1}]{Ashton}{Graves}
\author[affiliation={2}]{Sefik}{Emre Eskimez}
\author[affiliation={1}]{Shyamnath}{Gollakota}

\address{$^1$University of Washington, $^2$Sesame AI}

\affiliation{}{University of Washington}{USA}
\affiliation{}{Sesame AI}{USA}

\email{malek@cs.washington.edu, graveash@uw.edu, eeskimez@sesame.com, gshyam@cs.washington.edu}


\begin{document}

\maketitle

\begin{abstract}
We present the first neural target speech extraction (TSE) system that uses human feedback for iterative refinement. Our approach allows users to mark specific  segments of the TSE output, generating an edit mask. The refinement system then improves  the marked sections while preserving unmarked regions. Since large-scale datasets of human-marked errors are difficult to collect, we generate synthetic datasets using various automated  masking functions and train  models on each. Evaluations show that models trained with noise power-based masking (in dBFS) and probabilistic thresholding perform best, aligning with human annotations. In a  study with 22 participants, users showed a  preference for refined outputs over baseline TSE. Our findings demonstrate that human-in-the-loop refinement is a promising approach for improving the performance of neural speech extraction. 

\end{abstract}

\section{Introduction}


Despite advancements in model architectures and training techniques~\cite{i_vector, d_vector,zmolikova,tfgridnet}, neural speech extraction remains an unsolved problem, with no approach achieving consistently robust performance. Target speech extraction (TSE) models struggle to extract the target speaker when speech overlaps, the enrollment signal differs in acoustic characteristics from the mixture, or interfering speakers have similar vocal traits~\cite{speakerbeam}. Unlike background noise separation, distinguishing between subtle differences in human voices is far more complex~\cite{10.1145/3613904.3642057}. Consequently, TSE models may make mistakes in certain segments or incorrectly identify the speaker in some or all parts of the output.

\begin{figure}[t!]
  \centering
  \includegraphics[width=\linewidth]{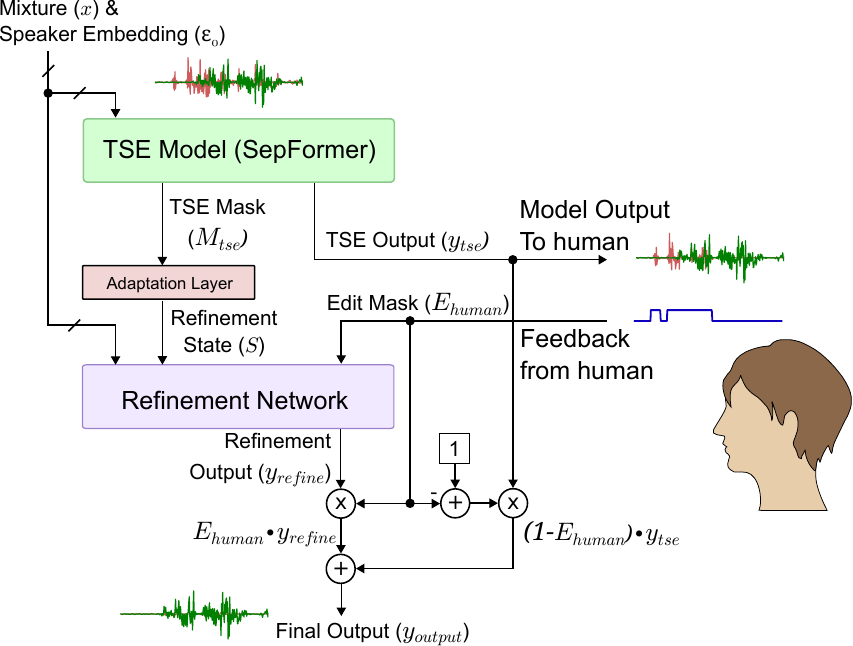}
  \vskip -0.1in
  \caption{System architecture for human-in-the-loop TSE. A neural TSE model extracts an initial target speech estimate, $y_{tse}$,  which a human reviews and marks for refinement. Our proposed refinement model then incorporates this human feedback to produce a more refined target speech estimate.}
  \vskip -0.2in
  \label{fig:architecture}
\end{figure}

Here, we present the first neural TSE system that incorporates human feedback for iterative refinement. As shown in Fig.~\ref{fig:architecture}, the system processes an input mixture using a TSE model to extract speech. If the output is unsatisfactory, the user can provide feedback  by marking specific segments  to generate an edit mask. Our refinement system  then utilizes the initial extraction and the edit mask to enhance the speech signal, modifying only the designated sections while preserving unaltered regions.

Human feedback has been helpful for aligning text generated by large language models with human preferences~\cite{bai2022traininghelpfulharmlessassistant} and has also been used to guide image editing to meet user needs~\cite{10204579,Ling2021EditGANHS}. However, building a refinement network for TSE using human feedback is challenging due to limited datasets.


Collecting large-scale datasets of human-marked errors in neural speech extraction output is not feasible. Instead, we create synthetic datasets that approximate human feedback. Specifically, we generate multiple synthetic datasets using various masking functions and train separate refinement models on each. These models are then evaluated on a set of 200 audio samples, annotated with edit masks from human annotators.  Our results show that the model trained with noise power-based masking function (in dBFS) with probabilistic thresholding yields the best improvements on  human-annotated samples. This aligns with human perception of loudness being roughly on a logarithmic scale~\cite{LITOVSKY201555}, while probabilistic thresholding accounts for variations in how humans create masks; making our synthetic data more representative of human feedback.


We develop an interactive tool that lets users listen to TSE model outputs and mark regions for refinement. To validate our system, we recruit 22 additional participants (10 annotators, 12 listeners) with varying audio  processing backgrounds. The annotators create edit masks for another 200 samples, and listeners rate the audio quality before and after refinement. Mean opinion scores reveal that participants prefer the refined audio from human feedback over the TSE-only output.



\section{Related work}


\noindent\textbf{Target speech extraction:} This task aims to extract a target speaker from a mixture using  cues such as audio examples \cite{zmolikova,i_vector,d_vector,speakerbeam,waveformer,10.1145/3613904.3642057}, spatial~\cite{hybridbeam, gu2019neural,acousticswarm}, visual~\cite{avsepformer}, text~\cite{liu2023separate}, or concept embeddings~\cite{conceptbeam}. While these works proposed various architectures to improve performance~\cite{tfgridnet,resepformer,li2024spmambastatespacemodelneed}, our work is complementary in that it addresses the imperfections of neural networks by integrating human feedback at inference time.

\vskip 0.05in\noindent\textbf{Audio editing.} Pre-deep learning audio editing tools, such as~\cite{10.1145/2556288.2557253}, enabled users to separate audio sources from a mixture by painting on time-frequency visualizations. More recent approaches employed transformers~\cite{lan2024musicongenrhythmchordcontrol} and diffusion models~\cite{10.5555/3692070.3694341} to enable modifications in both audio mixtures and music~\cite{morrison2024finegrainedinterpretableneuralspeech}. These methods leveraged text-based~\cite{ijcai2024p864} and instruction-guided methods~\cite{10.5555/3666122.3669246} to enable precise control over musical features like chords and rhythm~\cite{lan2024musicongenrhythmchordcontrol} as well as replaced audio classes in mixtures~\cite{10.5555/3666122.3669246}. More recent work~\cite{morrison2024finegrainedinterpretableneuralspeech} demonstrated editing of specific audio features, including speaker pitch, duration, volume, and spectral balance. However, none of these approaches addressed the challenge of multiple speech sources in a mixture or the task of target speech extraction.




\vskip 0.05in\noindent\textbf{Dynamic inference.} Prior work explored dynamic inference for speech separation using purely computational strategies~\cite{10638203}. Slimmable neural networks adjust the width of the network at run-time~\cite{10638203,10248143} while early exit methods~\cite{9413933,10096897} halt computation based on prediction similarity or gating decisions. In contrast, our approach integrates human feedback into neural speech extraction, allowing users to provide edit  masks  to refine the quality of the speech extraction. Our human-in-the-loop approach is complementary to existing deep learning-only strategies, enabling more accurate outputs.

\section{Methods}\label{Methods}

{\bf Problem Formulation.} Let $x\in\mathbb{R}^{T}$ be a noisy recording containing speech from a target speaker $s_0$, mixed with  $K$ interfering speakers $s_i\ (i=1,\dots,K)$ and  noise $n$. 
\begin{equation}
    x = s_{0} + \sum_{i=1}^{K}s_i + n
\end{equation}
The goal of neural TSE network, $\mathcal{F}$, is to extract an estimate ${s}^{tse}_0$ of the target speech  from $x$, using a speaker embedding $\varepsilon_{0}$:
\begin{equation}
y_{tse}={s}^{tse}_0 = \mathcal{F}(x|\varepsilon_{0})
\end{equation}
Since $\mathcal{F}$ may not always produce an accurate estimate of the target speech,  we seek to design a refinement network $\mathcal{G}$ that uses human feedback about the TSE output, ${s}^{tse}_0$. The feedback can be provided  in the form of a binary edit mask,  $E_{human}\in~\mathbb{Z}_2^T$, where $E_{human}[i] = 1$ if the user marks the $i$-th sample for refinement and $E_{human}[i] = 0$ otherwise. The goal  is to  obtain a refined estimate ${s}^{refined}_0$ that better approximates $s_0$:
\begin{equation}
y_{refine}={s}^{refined}_0 = \mathcal{G}(x|\varepsilon_{0};E_{human}; {s}^{tse}_0)
\end{equation}





\begin{figure}[t!]
  \centering
  \includegraphics[width=0.5\linewidth]{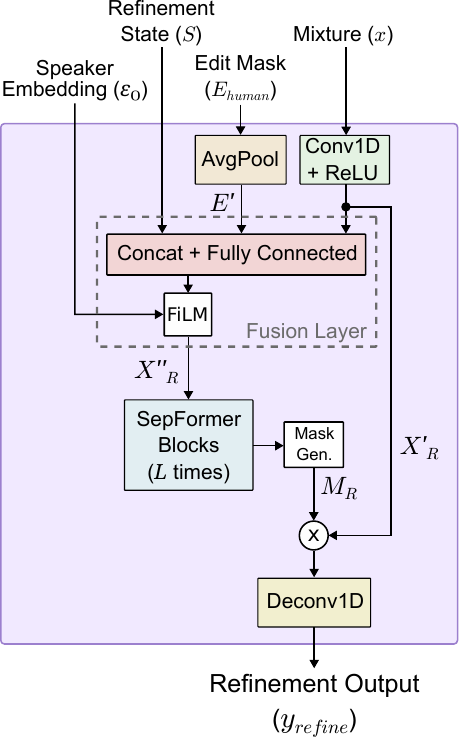}
  \vskip -0.1in
  \caption{Refinement network architecture. The encoded mixture, downsampled edit mask, TSE refinement state, and speaker embedding are fused into a conditioned tensor and processed by SepFormer blocks and a mask generator to produce a mask. }
  \vskip -0.2in
  \label{fig:refinearchitecture}
\end{figure}




\noindent{\bf System Architecture.}  
Fig.~\ref{fig:architecture} shows our human-in-the-loop neural TSE system.  The TSE network is based on SepFormer~\cite{sepformer}, a transformer-based architecture. We condition the model on the target speaker using a FiLM~\cite{film} layer, which is applied after the SepFormer encoder. We use d-vectors~\cite{d_vector} to condition the network on the target speaker characteristics. 

Formally, the TSE model encodes $x$ into a latent representation with $C_{tse}$ channels and $T'$ time steps. It then generates a mask $M_{tse}\in \mathbb{R}^{C_{tse} \times T'}$, which is applied to the encoded audio representation.   A learned decoder reconstructs the estimated time-domain target speech signal, $y_{tse} \in \mathbb{R}^{T}$. 

The output of the TSE network is presented to the  user, who provides feedback by marking samples that need refinement. 
Once the user provides the edit mask, $M_{tse}$ is transformed into a refinement state $S\in \mathbb{R}^{C_{R}\times T'}$ using a fully-connected adaptation layer, where $C_{R}$ is the number of refinement channels.  The refinement network then incorporates $S$ and the edit mask to generate the refined target speech output $y_{refine} \in {\mathbb{R}^{T}}$. We obtain the final estimate $y_{output} \in {\mathbb{R}^{T}}$ by updating only the sections  marked for refinement in the edit mask:
\begin{equation}
    y_{output} = E_{human} \cdot y_{refine} + (1 - E_{human}) \cdot y_{tse}
\end{equation}
\noindent{\bf Refinement Network.} 
Fig.~\ref{fig:refinearchitecture} shows how our network integrates human feedback to generate a more accurate approximation of the target speech. Similar to the TSE model, we first encode the mixture $x$ into a latent representation $X'_R \in \mathbb{R}^{C_{R}\times T'}$ using a strided convolution with a ReLU activation. This encoder doesn't share parameters with the TSE model.

To align the edit mask $E_{human}$ with the temporal resolution of $X'_R$, we apply average pooling to obtain a downsampled tensor $E'$ with the same number of time steps as $X'_R$. Then, we fuse $X'_R, E',$ the refinement state $S$, and $\varepsilon_0$ with a fusion layer. This layer performs channel concatenation, followed by a fully-connected layer, and finally, a FiLM layer to condition the input on the edit mask, refinement state, and speaker embedding. This fusion process produces a conditioned tensor $X''_{R}\in\mathbb{R}^{C_R \times T'}$. 

We then pass $X''_{R}$ through a series of $L$ SepFormer blocks, followed by a fully-connected mask generator to produce a refinement mask $M_R \in \mathbb{R}^{C_{R}\times T'}$. Finally, we multiply this mask with $X'_{R}$ and use a deconvolution layer to decode the refined target speech output $y_{refine} \in \mathbb{R}^{T}$.


\vskip 0.05in\noindent{\bf Automated Masking Functions as Substitutes for Human-Generated Masks.} To learn $\mathcal{G}$ using a data-driven approach, we require edit masks that capture human evaluations of TSE model outputs. However, collecting a sufficiently large dataset of human-annotated edit masks is impractical. Instead, we approximate them using masking functions.

Since the clean target speech $s_0$ represents an ideal reference, we assume that humans perceive deviations from $s_0$ in $s_0^{tse}$ as noise. When this deviation surpasses a certain threshold, a user would likely mark the region. Thus, our masking function quantifies the dissimilarity between $s_0$ and $s_0^{tse}$, assigning ones to regions exceeding the threshold and zeros otherwise.

Formally, we define the masking function as a mapping $f(A, B):\mathbb{R}^{N} \times \mathbb{R}^{N} \to \mathbb{Z}_2^{N}$, where the synthetic edit mask is computed as $E_{synthetic} = f(s_0^{tse}, s_0)$. Our goal is to choose a masking function such that $E_{synthetic}$ closely aligns with human annotations, i.e., $E_{synthetic} \sim E_{human}$.

To compute the edit mask, we segment $s_0$ and $s_0^{tse}$ into non-overlapping 0.25-second windows, i.e., $N=4000$ at a sampling rate of 16~kHz, and  calculate the masking function between pairs of windows corresponding to the same segment in time, and concatenate the results. This produces a fine-grained edit mask that varies over time. Additionally, we also define a global edit mask, which applies a single value across the entire signal, either marking all or none of it for refinement.

Masking functions have the following form:

$f(A, B) = \begin{cases}
\mathbf{1}^N & \text{if } g(A, B) > \tau\\
\mathbf{0}^N & \text{otherwise}
\end{cases}$\\
 
\noindent where $g:\mathbb{R}^{N} \times \mathbb{R}^{N} \to \mathbb{R}$ is a similarity metric and $\tau$ is a threshold value. In this work, we look at five different masking functions, which differ in the choice of $g$ and $\tau$:
 

\begin{itemize}
    \item Fine-grained Mean Absolute Error (\textit{meanAE}):\\
    $g(A, B) = \frac{1}{N}\sum_{i}|A_i - B_i|;\ \tau = 0.03$
    
    \item Fine-grained Max Absolute Error (\textit{maxAE}): \\
    $g(A, B) = \text{max}_i|A_i - B_i|;\ \tau = 0.1$
    
    \item Fine-grained decibels relative to full-scale (\textit{dBFS}): \\
    $g(A, B) = 10 \log\frac{1}{N}\sum_{i}(A_i - B_i)^2;\ \tau = -40$
    
    \item Fine-grained dBFS, probabilistic (\textit{dBFS-prob}): \\
    $g(A, B) = 10 \log\frac{1}{N}\sum_{i}(A_i - B_i)^2;\ \tau \sim \mathcal{N}(-40, \sigma=3)$
    
    \item Global signal-to-noise ratio (\textit{GlobalSNR}): \\
    $g(A, B) = -\text{SNR}(A, B);\ \tau = -5$
\end{itemize}

\noindent $\text{SNR}(\hat{x}, x) = 10\log_{10}\big(\frac{||x||_2^2}{||x - \hat{x}||_2^2}\big)$ is the signal-to-noise ratio.

\begin{table}[t]
  \caption{Comparing model configurations and masking functions. TSE+Refine is our proposed refinement strategy, while TSE+TSE  successively applies the same TSE model whenever a masking function has any non-zero sample.}
  \vskip -0.1in
  \label{tab:main_results}
  \centering
  \setlength{\tabcolsep}{2.3pt}

  \begin{tabular}{ c c c c c c c c c }
    \toprule
    \multicolumn{1}{c}{\textbf{Config.}} &
    \multicolumn{1}{c}{\textbf{Masking}} &
    \multicolumn{1}{c}{\textbf{Count}} & \multicolumn{1}{c}{\textbf{SISDR}} & \multicolumn{1}{c}{\textbf{PESQ}} & \multicolumn{1}{c}{\textbf{DNSMOS}}  \\
    
    & \textbf{function} & & \multicolumn{1}{c}{\textbf{(dB)}} & & \multicolumn{1}{c}{\textbf{OVRL}}\\

    \midrule

    Mixture & -- & -- & 0.01 & 1.26 & 2.48 \\
    
    \midrule
    
    TSE & -- & -- & 12.18 & 1.91 & 3.21 \\

    \midrule

    TSE+Refine & meanAE & 109 & 12.92 & 1.92 & 3.22 \\
    TSE+TSE & meanAE & 109 & 11.92 & 1.91 & 3.21 \\

    \midrule
    
    TSE+Refine & maxAE & 798 & 14.03 & 2.03 & 3.37 \\
    TSE+TSE& maxAE & 798 & 9.72 & 1.80 & 3.12 \\
    
    \midrule
    
    TSE+Refine & GlobalSNR & 41 & 12.72 & 1.93 & 3.23 \\
    TSE+TSE & GlobalSNR & 41 & 12.09 & 1.91 & 3.21 \\

    \midrule
    
    TSE+Refine & dBFS & 945 & 14.07 & 2.07 & 3.40 \\
    TSE+TSE & dBFS & 945 & 9.18 & 1.76 & 3.08 \\

    \midrule
    
   TSE+Refine &  dBFS-prob & 917 & \textbf{14.88} & \textbf{2.16} & \textbf{3.49} \\
   TSE+TSE & dBFS-prob & 917 & 9.25 & 1.77 & 3.09 \\
     
    \bottomrule
  \end{tabular}
      \vskip -0.1in
\end{table}

\begin{table}[t]
  \caption{TSE+Refinement results using different masking functions with 2-speaker VCTK mixtures using human annotations.}
  \vskip -0.1in
  \label{tab:human_annotations}
  \centering
  \begin{tabular}{ c c c c }
    \toprule
    \multicolumn{1}{c}{\textbf{Config/}} &
    \multicolumn{1}{c}{\textbf{SI-SDR}} & \multicolumn{1}{c}{\textbf{PESQ}} & \multicolumn{1}{c}{\textbf{DNSMOS}}\\
    \multicolumn{1}{c}{\textbf{Masking function}} & \multicolumn{1}{c}{\textbf{(dB)}} & &\multicolumn{1}{c}{\textbf{OVRL}}\\

    \midrule

    
    TSE & -0.93 & 1.60 & 2.72 \\
    
    \midrule
   {\bf TSE+Refine} & & &\\
    meanAE & 2.40 & 1.72 & 3.05 \\
    maxAE & 4.75 & 1.83 & 3.00 \\
    GlobalSNR & -5.73 & 1.61 & \textbf{3.07}\\
    dBFS & 4.57 & 1.80 & 2.99\\
    dBFS-prob & \textbf{5.76} & \textbf{1.85} & 3.02\\
     
    \bottomrule
  \end{tabular}
      \vskip -0.2in
\end{table}

\begin{table}[t]
  \caption{Evaluation on 2-speaker VCTK mixtures.}
  \vskip -0.1in
  \label{tab:vctk}
  \centering
  \begin{tabular}{ c c c c }
    \toprule
    \multicolumn{1}{c}{\textbf{Config.}} &
    \multicolumn{1}{c}{\textbf{SI-SDR (dB)}} & \multicolumn{1}{c}{\textbf{PESQ}} & \multicolumn{1}{c}{\textbf{DNSMOS}}\\

    \midrule

    Mixture & -0.16 & 1.54 & 2.98 \\
    TSE & 10.81 & 2.02 & 3.07 \\
   TSE+Refine & \textbf{13.08} & \textbf{2.22} & \textbf{3.28}\\
     
    \bottomrule
  \end{tabular}
      \vskip -0.1in
\end{table}

\begin{table}[t]
  \caption{Evaluation on 3-speaker LibriSpeech mixtures.}
  \vskip -0.1in
  \label{tab:3spk}
  \centering
  \begin{tabular}{ c c c c }
    \toprule
    \multicolumn{1}{c}{\textbf{Config.}} &
    \multicolumn{1}{c}{\textbf{SI-SDR (dB)}} & \multicolumn{1}{c}{\textbf{PESQ}} & \multicolumn{1}{c}{\textbf{DNSMOS}}\\

    \midrule

    Mixture & -0.3 & 1.16 & 2.14 \\
    TSE & 8.21 & 1.51 & 2.81 \\
    TSE+Refine & \textbf{10.62} & \textbf{1.65} & \textbf{3.06}\\
     
    \bottomrule
  \end{tabular}
      \vskip -0.1in
\end{table}

\begin{table}[t]
  \caption{Evaluation on noisy 2-speaker LibriSpeech mixtures.}
  \vskip -0.1in
  \label{tab:noisy}
  \centering
  \begin{tabular}{ c c c c }
    \toprule
    \multicolumn{1}{c}{\textbf{Config.}} &
    \multicolumn{1}{c}{\textbf{SI-SDR (dB)}} & \multicolumn{1}{c}{\textbf{PESQ}} & \multicolumn{1}{c}{\textbf{DNSMOS}}\\

    \midrule

    Mixture & -0.15 & 1.17 & 2.37 \\
    TSE & 10.48 & 1.61 & 2.96 \\
    TSE+Refine & \textbf{12.27} & \textbf{1.72} & \textbf{3.22}\\
     
    \bottomrule
  \end{tabular}
      \vskip -0.2in
\end{table}

\section{Experiments and Results}

{\bf Datasets.}  We trained our models on 16 kHz~speech from LibriSpeech~\cite{librispeech} and noise from WHAM!~\cite{wham}, generating training mixtures on-the-fly via dynamic mixing. Each 5-second mixture was created by randomly selecting $K$ speaker utterances from the same corpus split. Utterances longer than 5 seconds were cropped, and shorter ones were zero-padded with random silence. One speaker was designated as the target, with their d-vector embedding derived from a separate utterance.

Each training epoch included 20,000 mixtures, with validation and test sets fixed at 2,000 and 1,000 samples, respectively. Speech data came from LibriSpeech’s \verb|train-clean-360|, \verb|test-clean|, and \verb|dev-clean| splits, while noise data was sampled from WHAM!’s \verb|tr|, \verb|cv|, and \verb|tt| splits. Interferer and noise amplitudes were scaled for a target speaker SNR uniformly distributed between -10 dB and 10 dB.


\vskip 0.03in\noindent\textbf{Training setup.} We first trained the TSE model independently, then froze its weights and trained the adaptation layer and refinement network together. All models were trained for 300 epochs. For TSE models, the learning rate (LR) started at 0.002, halving after 4 epochs of no validation loss improvement. For refinement models, LR started at 0.001, halving after 6 stagnant epochs. All models used the AdamW optimizer with weight decay 0.01 and a gradient clipping of 1. The TSE model was based on SepFormer, with an encoder using a kernel size of 32 and output channel dimension $C_{tse} = 64$.  SepFormer had a chunk size of 250, $L=2$ layers, and intra-/inter-attention modules with 8 attention heads and 4 repetitions. The refinement model had the same configuration, with $C_R =  64$. The models were trained to minimize negative SI-SDR~\cite{sisdr} between outputs and ground truth speech. While metrics were computed on the final output $y_{output}$ during evaluation, training and validation losses were based on the refinement model output $y_{refine}$. Results are reported using the model weights with the lowest validation loss.

\vskip 0.05in\noindent{\bf Refinement versus Successive TSE.} 
We evaluate our refinement strategy using SI-SDR~\cite{sisdr}, PESQ~\cite{pesq}, and personalized DNSMOS~\cite{dnsmos} across all input samples. We first assess the system by training and testing with different masking functions, measuring speech quality  after refinement. Here, we focus on the two-speaker case without background noise. We also evaluate a baseline approach where samples needing refinement are  passed again through the original TSE network.

Table~\ref{tab:main_results} shows that repeatedly applying the TSE network degrades performance, whereas the proposed refinement method  enhances target speech quality across all evaluation metrics and masking functions. This degradation likely stems from the TSE model over-suppressing the target speaker, leading to irreversible information loss in the absence of the original mixture. Among the tested masking functions, dBFS-prob achieves the highest performance gains, improving SI-SDR, PESQ, and DNSMOS OVRL from 12.18 dB, 1.91, and 3.21 (TSE only) to 14.88 dB, 2.16, and 3.49, respectively.

The \texttt{count} variable represents the number of TSE output samples identified for refinement under each masking function. Unlike GlobalSNR, which flags only samples with an average output SNR below 5 dB over the full 5-second recording, dBFS-prob exhibits greater sensitivity to localized errors, leading to a higher number of samples selected for refinement.




\vskip 0.05in\noindent{\bf Validating masking functions with human annotations.} To this end, we first generated 200 TSE output samples, as our refinement procedure operates on TSE outputs. These TSE output samples were selected to ensure a uniform distribution of SI-SDR {with an average SI-SDR close to  0 dB.}

\begin{figure}[t]
  \centering
    \caption{Interactive tool interface used for human evaluation.}
    \vskip -0.15in
\includegraphics[width=0.61\linewidth]{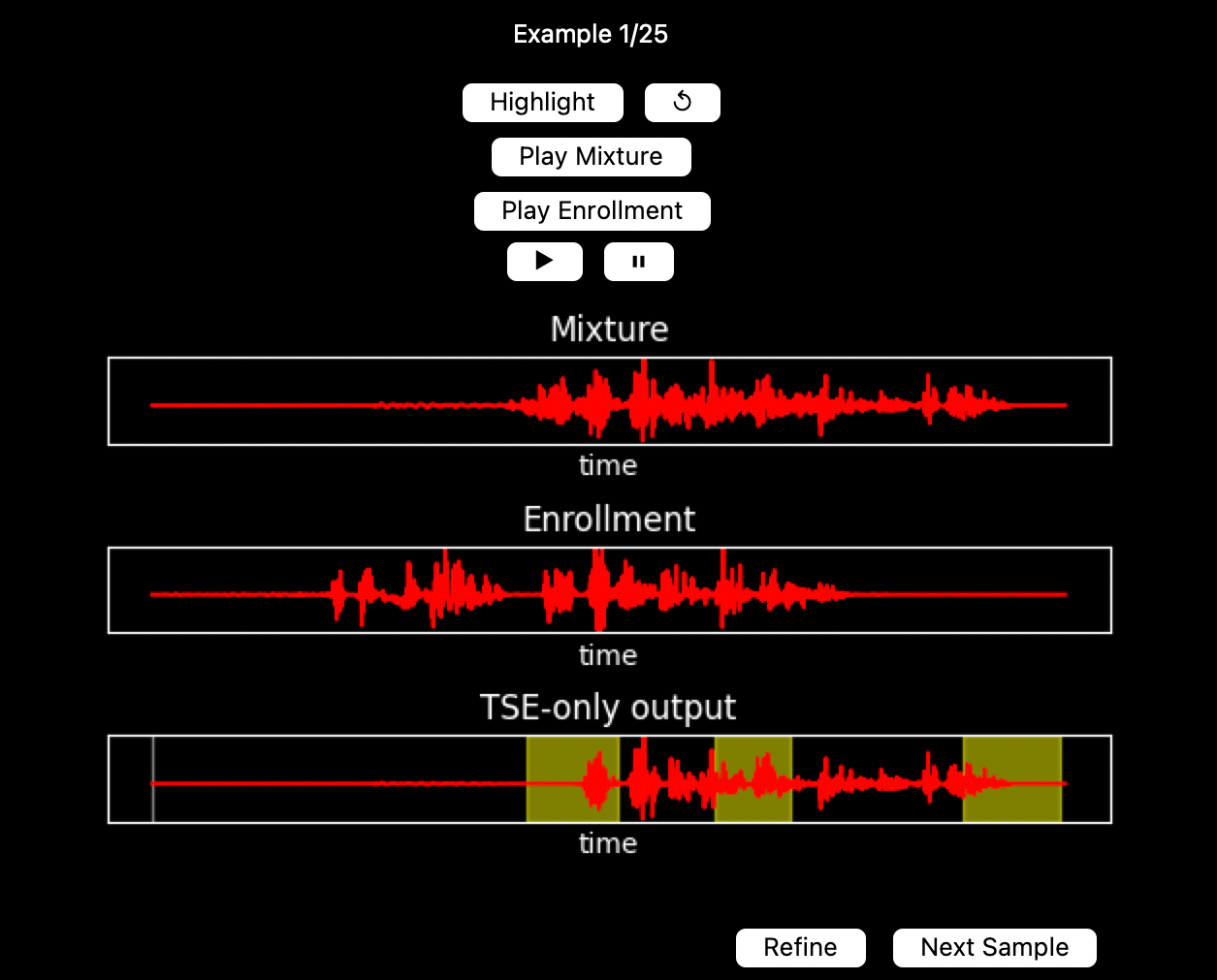}
\vskip -0.1in
  \label{fig:interface}
\end{figure}

We developed an interactive tool (Fig.~\ref{fig:interface}) that enables users to listen to and visualize the time-domain waveforms of the mixture, enrollment audio, and TSE output. Four participants were recruited to use this tool to annotate regions in the TSE output that required refinement. These annotations were then converted into edit masks, where samples were assigned a value of 1 in the marked regions and 0 elsewhere.

After collecting the human edit masks, we applied our refinement models using these masks. The results in Table~\ref{tab:human_annotations} show that our refinement method  improves speech quality. However, the choice of masking function  influences its transferability to human annotations. The dBFS-prob masking function yields the best performance, improving SI-SDR by 6.69 dB over the TSE output. Additionally, it increases SI-SDR by approximately 1.2 dB over dBFS, suggesting that a probabilistic threshold can serve as an effective data augmentation strategy to capture diverse user preferences. For all subsequent evaluations, we use the model trained with dBFS-prob as the default. 



\vskip 0.05in\noindent{\bf Additional experiments.} To evaluate our models on a different dataset,  we created a test set of 1,000 two-speaker speech mixtures following the same procedure as before but using speech data from the VCTK corpus~\cite{vctk}. These mixtures were processed using our TSE refinement strategy trained on LibriSpeech mixtures. As shown in Table~\ref{tab:vctk}, our method improves the average SI-SDR over the TSE model by 2.27~dB, demonstrating its ability to generalize to out-of-distribution datasets.

We also evaluate our refinement strategy on datasets with three speakers and two speakers plus noise. For each of these scenarios, we train separate TSE and refinement models. The results are shown in Tables~\ref{tab:3spk} and \ref{tab:noisy}, respectively. Our refinement algorithm consistently improves all metrics in multiple background speakers and noisy scenarios. In the 3-speaker case, refinement improves the SI-SDR by 2.41~dB on average, while in the noisy speaker case, it can improve by 1.79~dB.

Finally, we replace average pooling with a 1D strided convolution (kernel size 32, stride 16, 1 output channel). Testing on the 2-speaker LibriSpeech dataset with the dBFS-prob masking function yields an SI-SDR difference within 0.06 dB.


\begin{table}[t]
  \caption{Objective and subjective results for human evaluation. Since refinement is useful only when TSE underperforms, samples were selected so the average TSE output 
SI-SDR is  $\sim$0 dB.}
  \vskip -0.1in
  \label{tab:survey}
  \centering
  \begin{tabular}{ c c c c }
    \toprule
    \multicolumn{1}{c}{\textbf{Config.}} &
    \multicolumn{1}{c}{\textbf{SI-SDR (dB)}} & \multicolumn{1}{c}{\textbf{PESQ}} & \multicolumn{1}{c}{\textbf{MOS}}\\

    \midrule

    TSE & -0.55 & 1.59 & 2.10\\
    TSE+Refine & 4.79 & 1.80 & \textbf{2.70} \\
    TSE+Refine-replace & \textbf{4.96} & \textbf{1.85} & 2.55\\
     
    \bottomrule
  \end{tabular}
      \vskip -0.2in
\end{table}

\vskip 0.05in\noindent{\bf Human subjective evaluation.} Finally, we evaluate our TSE refinement system using human-annotated edit masks from a completely new set of annotators. We created a new  dataset of 200 mixtures, ensuring that the SI-SDR distribution aligned with that of the previous human evaluation. Ten additional random annotators used our tool to listen to and annotate regions for refinement. Each participant annotated 25 samples, with the first five serving as a familiarization phase and subsequently discarded. The remaining 20 annotations per participant were converted into edit masks using the same procedure described earlier. We applied our refinement algorithm to these 200 human-annotated samples and computed the objective results, shown in Table~\ref{tab:survey}. Our method consistently improves SI-SDR and PESQ, demonstrating that both our approach and the selected masking function generalize effectively to unseen annotators.

To assess subjective quality, we recruited 12 additional participants to rate the quality of the  the TSE output, and our refined output for a randomly selected 15 audio examples from the 200 annotated samples. These samples were presented with an enrollment audio of the target speaker. Table~\ref{tab:survey} shows that participants  favored our refined output, with the mean opinion score (MOS) increasing by 0.6 points. {Paired t-tests between our TSE+Refine model and TSE, and between TSE+Refine model and TSE+Refine-replace show a statistically significant difference with p $\lt$ 0.01 and p $\lt$ 0.1 respectively.} This confirms that our refinement system enhances both objective speech quality and human-perceived audio clarity. {Interestingly, while  TSE+Refine-replace, which uses $y_{refine}$ and not $y_{output}$,  improves  objective metrics, the participants preferred  TSE+Refine. This may be due to the refinement model introducing subtle artifacts outside the annotated regions,  impacting how user perceive the overall audio quality.}

\section{Conclusion}
We present a neural speech extraction system incorporating human feedback for iterative refinement. Our work has limitations offering exciting future research opportunities. We are focused on a single refinement iteration to minimize user effort. Exploring multi-iteration refinement networks improving performance while minimizing user effort is valuable. Our system allows marking segments for refinement, but detailed within-segment feedback (e.g., ``reduce noise further'') could be explored. Finally, exploring generative models, like diffusion models, with human feedback for TSE could yield additional improvements.

\noindent{\bf Acknowledgments.} The researchers are partly supported
by the Moore Inventor Fellow award \#10617, Thomas J. Cable Endowed Professorship, and a UW CoMotion innovation gap fund. This work was facilitated through the use of computational, storage, and networking infrastructure provided by the UW HYAK Consortium.

\bibliographystyle{IEEEtran}
\bibliography{paper}

\end{document}